# Multiferroic Quantum Phase Transitions in Uniaxial Hexaferrites BaFe$_{12}$O$_{19}$, SrFe$_{12}$O$_{19}$, and PbFe$_{12-x}$Ga$_x$O$_{19}$

S. E. Rowley,[1,2] Yi-Sheng Chai,[3] Shi-Peng Shen,[3] Young Sun,[3] A. T. Jones,[1] B. E. Watts,[4] and J. F. Scott[1,5]

[1]Cavendish Laboratory, Cambridge University, J. J. Thomson Avenue, Cambridge, CB3 0HE, United Kingdom
[2]Centro Brasileiro de Pesquisas Físicas, Rua Dr Xavier Sigaud 150, Urca, Rio de Janeiro, 22290-180, Brazil
[3]Beijing National Laboratory for Condensed Matter Physics, Institute of Physics, Chinese Academy of Sciences, Beijing 100190, China
[4]IMEM-CNR, Parco Area delle Scienze 37/A, 43124 Parma, Italy
[5]Schools of Chemistry and of Physics and Astronomy, St. Andrews University, St. Andrews, Fife, KY16 9ST, United Kingdom

BaFe$_{12}$O$_{19}$ is a popular M-type hexaferrite with a Néel temperature of 720 K and is of enormous commercial value ($3 billion/year). It is an incipient ferroelectric with an expected ferroelectric phase transition extrapolated to lie at 6 K but suppressed due to quantum fluctuations (as in SrTiO$_3$). The theory of quantum criticality for such uniaxial ferroelectrics predicts that the temperature dependence of the electric susceptibility χ diverges as $1/T^3$, in contrast to the $1/T^2$ dependence found in pseudo-cubic materials such as SrTiO$_3$ or KTaO$_3$. In this paper we present evidence of the susceptibility varying as $1/T^3$, i.e. a critical exponent γ = 3. In general γ = (d + z − 2)/z, where the dynamical exponent for a ferroelectric z = 1 and the effective dimension is increased from d = 3 to d = 4 due to the effect of long-range dipole interactions in uniaxial as opposed to multiaxial ferroelectrics. The electric susceptibility of the incipient ferroelectric SrFe$_{12}$O$_{19}$, which is slightly further from the quantum phase transition is also found to vary as $1/T^3$ up to around 35K. Finally, by replacing Fe with approximately 75% Ga, the Néel temperature of 720K in isomorphic PbFe$_3$Ga$_9$O$_{19}$ is suppressed to absolute zero, resulting in a magnetic quantum phase transition.

Hexagonal ferrites are the most common magnetic materials with 90% of the $4 billion world market. 300,000 tons of hexagonal BaFe$_{12}$O$_{19}$ are produced every year, which corresponds to 50 grams for every person on Earth[1,2]. Primary uses are

magnetic credit cards, bar codes, and small motors, as well as low-loss cheap microwave devices. In 2011 Fujifilm produced a barium hexaferrite-based tape with a memory of five terabytes − the equivalent of eight million books. At present this material has a new aspect of fundamental interest − it is nearly ferroelectric as the temperature approaches absolute zero. Incipient ferroelectrics at low temperature, i.e. materials close to a ferroelectric quantum phase transition, are expected to be important for a wide range of advanced applications including for example, electro-caloric refrigeration, quantum memory devices, and cryogenic electronic switches, as their properties can be readily controlled by voltage gates and strains.

Very recently we examined[3] the quantum criticality of uniaxial ferroelectric tris-sarcosine calcium chloride-bromide (TSCC:Br) and found that the low-temperature dielectric constant diverged with temperature as $1/T^2$, as in pseudo-cubic compounds such as strontium titanate[4] and in contrast to the inverse cubic dependence first predicted by Khmelnitskii and Shneerson[5], and were able to show that this paradox arises from the ultra-weak ferroelectric dipoles in that material. Here we report the study of a second uniaxial paraelectric, M-type barium hexaferrite, near its ferroelectric quantum phase transition, which is a strongly displacive system, with an $A_{2u}$ symmetry soft mode frequency decreasing to 42 cm$^{-1}$ as $T$ goes to zero[6]. The only other low-$T$ multiferroic studied in detail previously is EuTiO$_3$[7] which is slightly too far from the critical point to manifest quantum critical behaviour.

There has been some controversy concerning ferroelectricity in this family of M-type hexaferrites: polarization-electric field hysteresis loops $P(E)$ of SrFe$_{12}$O$_{19}$ at 300 K were published by Tan and Wang[8,9], and there is also a recent theoretical paper[10] by Wang and Xiang that predicts a paraelectric to antiferroelectric phase transition for BaFe$_{12}$O$_{19}$ at about 3.0 K. In this context it is important to note that SrFe$_{12}$O$_{19}$ and (Ba,Sr)Fe$_{12}$O$_{19}$ are n-type semiconductors[11] with bandgaps at approximately $E_g = 0.63$ eV and rather heavy electrons and holes: m(light e) = 5.4 m$_e$; m(heavy e) = 15.9 m$_e$; m(light h) = 10.2 m$_e$; m(heavy h) = 36.2 m$_e$ and highly anisotropic conductivity, so it is important to discriminate between true ferroelectric hysteresis and leakage current artefacts. For electric fields applied normal to the c-axis, the electrical conductivity is ca. 50× greater than along c, which will create strong leakage currents. The present work and ref. 12 show that these suggested ferroelectric transitions do not occur at finite temperatures and that $BaFe_{12}O_{19}$ retains its paraelectric $P6_3/mmc$ symmetry ($D_{6h}$) down to zero temperature. When fitting the inverse susceptibility $1/\chi_E$ to a Curie-Weiss law at higher temperatures (the linear part of the curve), an extrapolation to $1/\chi_E = 0$ gives an expected Curie temperature, like that in $SrTiO_3$ or $KTaO_3$, at ca. 6 K (similar to the 35 K value in $SrTiO_3$). However the anticipated ferroelectric state does not stabilize and is suppressed by quantum fluctuations (as in the freezing temperature of liquid helium) resulting in a paraelectric ground state with quantum critical fluctuations. The proximity to the quantum critical point is evident from a rapidly rising dielectric susceptibility as the temperature is lowered and a soft $A_{2u}$-symmetry $q=0$ long wavelength phonon mode that decreases to 42 cm$^{-1}$ as $T$ approaches zero[6]; we designate this minimum (gap) in the transverse-optical phonon frequency in the low temperature limit as $\Delta$. It is this Curie-Weiss extrapolated transition at 6K that produced a fictitious 3K specific heat anomaly in the model of Wang and Xiang[10]. A good review of work on $SrFe_{12}O_{19}$ was given this year by Hilczer et al.[13] From a magnetic point of view Ba-hexaferrite is unusual in that although all 24 spins per primitive unit cell (two formula groups) are $Fe^{+3}$, it is a ferrimagnet with 8 spins up (at tetrahedral, octahedral, and five-fold coordinated sites) and 16 down (all at octahedral sites), as shown in Fig. 1a. This produces a strong ferromagnetic moment, unlike weak canted antiferromagnets (it is often termed a Lieb-Mattis ferrimagnet[14]).

M-type hexaferrite single crystals were prepared by the flux method. The single-crystal x-ray diffraction (XRD) patterns at room temperature shown in Fig. 1c suggest that our samples are single-phase M-type with $c = 23.18$ Å for $BaFe_{12}O_{19}$ and 23.04 Å for $SrFe_{12}O_{19}$, respectively, agreeing with the original 1959 single-crystal value of Brixner[15,16]. The structure is not completely agreed upon in the literature:

Ganapathi et al.[17] report a tripled unit cell along the a-axis for flux-grown crystals like those in the present work; this differs from the original structural determination[15] with $a = 5.895$ Å. However, this is unimportant for the present study since the ferroelectric properties are thought to involve only symmetry changes along the c-axis.

In a previously published paper[12] we found that M-type ferrimagnetic hexaferrites (Ba,Sr)$Fe_{12}O_{19}$ are a new family of magnetic quantum paraelectrics along the *c*-axis only. This preservation of c-axis six-fold symmetry is compatible with the $A_{2u}$-symmetry soft mode reported from the Rostov group, which retains the hexagonal symmetry. The resulting symmetry of the crystal, were it to undergo a transition into a ferroelectric phase, is therefore probably $C_{6v}$ point group symmetry and $P6_3/mc$ space group. Because there is no change in hexagonal crystal class, this transition would be purely ferroelectric and not ferroelastic[18], with no hysteresis in its stress-strain relationship. That may be important with regard to descriptions of the system close to quantum criticality, implying that no elastic order parameter is a conjugate force.

As shown in Fig. 1b, M-type hexaferrite exhibits a new mechanism for local electric dipoles based on the magnetic $Fe^{3+}$ ($3d^5$) ion, violating the $d^0$ rule of Nicola Hill[19]. The competition between the long-range Coulomb interaction and short-range Pauli repulsion in a $FeO_5$ bipyramid with proper lattice parameters would favour an off-centre displacement of $Fe^{3+}$ that induces a local electric dipole. Such local dipoles cannot order down to the lowest temperatures in the specimens we measured but ferroelectric ground states may be reached perhaps via tuning with strains, chemical substitution or by varying the lattice density.

**Dielectric Measurements**

Our low temperature inverse electric susceptibility data $1/\chi_E$, (related to the measured dielectric constant $\varepsilon$ by $\chi_E = \varepsilon - 1$), are shown in Fig. 2. The measurements were obtained with a pumped helium-3 cryostat for both cooling and heating cycles, typically at rates of 10 mK/minute (overnight runs). Here we see that below ca. $T = 6K$ in $BaFe_{12}O_{19}$ and 20K in $SrFe_{12}O_{19}$ there is a non-monotonic dependence; we have previously reported such effects in $SrTiO_3$, $KTaO_3$[4], and tris-sarcosine calcium chloride (TSCC) and shown quantitatively without adjustable parameters in the former cases that they arise from acoustic phonon coupling (electrostriction). In those cases, the upturn in the inverse susceptibly as determined by measurements and theory without adjustable parameters, occurs when $T$ is less than 10% of $T_x$ where $T_x$ is the temperature scale associated with soft transverse-optical phonon frequency at the zone centre $\Delta$ in the zero temperature limit, i.e. $T_x = \hbar\Delta/k_B$. This means that we can attempt to fit the dielectric susceptibility data only for $T > 0.1T_x$ to a quantum criticality model in the absence of magneto-electric or electro-strictive coupling terms, the parameters for which are not currently available for our samples. $0.1T_x = 6K$ for $BaFe_{12}O_{19}$ as determined from measurements[6], and estimated to be 20K in $SrFe_{12}O_{19}$ which is further away from the quantum phase transition. Note that precisely at a ferroelectric quantum critical point, the frequency gap $\Delta$ vanishes and both $T_x$ and the Curie temperature $T_C$ are exactly zero. This means that such upturns only exist in samples with paraelectric ground states some distance away, but close to, the quantum phase transition. Another cross-over temperature exists for the upper temperature limit for any power-law exponent: In measurements and theory in $SrTiO_3$ and $KTaO_3$ we found that a single quantum critical exponent extends up to ca. 10% of the characteristic temperature $T^*$. $T^*$ is analogous to the Debye temperature but of the soft (critical) transverse-optical phonon mode and is given by $T^* = \hbar vQ/k_B$ where $v$ is the gradient of the frequency with respect to wavevector $q$ at low temperatures from a dispersion of the form $\Omega^2 = \Delta^2 + v^2q^2 + \cdots$ in the limit that $\Delta$ goes to zero, and $Q$ is the value of $q$ at the Brillouin zone boundary. It can also be estimated from $k_BT^* = \hbar\Omega(q = Q)$ where $\Omega(q = Q)$ is the value of the transverse-optic mode frequency measured in the low temperature limit at the Brillouin zone boundary, $q=Q$. The soft mode dispersion in $BaFe_{12}O_{19}$ has not been measured, but based upon its frequency at $q=0$ and the heavy masses in $BaFe_{12}O_{19}$, we estimate $T^*$ to be approximately 150K. Therefore in Fig. 2c we fit the measured dielectric susceptibility $\chi_E(T)$ from about 6-15 K for $BaFe_{12}O_{19}$, i.e. between the lower and upper crossover temperatures $0.1T_x$ and $0.1T^*$ respectively. Although this temperature interval is small, Fig. 2c shows that $1/\chi_E$ varies as $T^3$ for ca. 700 data points in the region, thus supporting the theory of Khmelnitskii and Shneerson[5] of quantum criticality for a uniaxial ferroelectric. Above 10% of $T^*$ one expects the system to exhibit classical Curie-Weiss behaviour as observed in our data in Figs. 2a and 2b. High pressure work by which $T_c$ may be brought up through $T = 0$ will be the subject of future work, but the results of dielectric measurements in the presence of "chemical pressure" obtained by replacing Ba-ions with smaller Sr-ions is shown in Fig. 2b and 2d, where a cubic temperature dependence agrees with the measured data in the range 20 K to 35 K for $SrFe_{12}O_{19}$.

**Other possible phase transitions**

In addition to the incipient ferroelectric phase near $T = 0K$, there is known to be a phase transition in $SrFe_{12}O_{19}$ (but not yet reported in $BaFe_{12}O_{19}$) near 55K at which the dynamic disorder of one set of Fe-ions between closely spaced sites becomes static[20,21]. The exact space group symmetry is not known with certainty, but it appears to be $D_{6h}$(disordered) to $D_{6h}$(ordered) or possibly $C_{6h}$. If tuned into the ferroelectric phase there would be a further symmetry lowering to $C_{6v}$. In addition, a magnetic phase transition has been reported very near 80K[21,22].

**Magnetic quantum phase transitions**

By replacing approximately 9 of the 12 of the $Fe^{+3}$ ions per formula group in M-type hexaferrites with Ga, it is possible to lower $T_N$ all the way to $T = 0$ K. Fig. 3a shows the dielectric behaviour in $PbFe_{12-x}Ga_xO_{19}$ with a low values of $T_N$. There is a small peak in the real part $\varepsilon'(T)$ near $T_{Néel} = 15K$, which unlike the $BaFe_{12}O_{19}$ and $SrFe_{12}O_{19}$ systems discussed above, arises from magnetoelectric coupling, presumably via striction, plus a large divergence in $\varepsilon''(T)$ near $T = 0$ K, which we interpret as arising from magnetocapacitance. A model for dielectric loss at Néel temperatures has been given by Pirc et al[28], and their graph of ε'(*T*) and ε''(*T*) for low-frequency probes is given in Fig. 3 of ref. 28 for realistic parameters, assuming an indirect magnetoelectric interaction through striction. The present data satisfy a Vogel-Fulcher relationship with frequency from 100 Hz to 1 MHz (Fig. 3b).

**Methods**

M-type hexaferrite single crystals were prepared by the flux method. The raw powders of $BaCO_3$ ($SrCO_3$), $Fe_2O_3$, and fluxing agent $Na_2CO_3$ were weighed in the molar ratio 10.53% : 26.3% : 63.17% and were well mixed. The mixed raw powder was put in a Pt crucible and heated to 1250 °C for 24h in air, then cooled down to 1100 °C at a rate of 3 °C/min and finally quenched to room temperature. The samples (ca. 2 mm on a side) were characterized by single-crystal x-ray diffraction at room temperature by using a Rigaku X-ray diffractometer.

The dielectric measurements were carried out in a pumped helium-3 cryostat at temperatures as low as 0.3 K and a liquid-cryogen-free superconducting magnet system (Oxford Instruments, TeslatronPT) down to 1.5 K. Silver paste was painted on the surfaces (ab plane) of a thin plate of each crystal and an Andeen-Hagerling or Agilent 4980A LCR meter was used to measure the dielectric susceptibility at frequencies typically in the range 1 kHz to 1 MHz.

**References**

[1] R. C. Pullar, Hexagonal ferrites: A review of the synthesis, properties and applications of hexaferrite ceramics, Prog. Mat. Sci. 57, 1191-1334 (2012).
[2] R. C. Pullar, Multiferroic and Magnetoelectric Hexagonal Ferrites, Springer Series in Materials Science (eds. A. Saxena and A. Planes): Volume 198, Mesoscopic Phenomena in Multifunctional Materials: Synthesis, Characterization, Modelling and

Applications (Heidelberg, 2014).

[3] S. E. Rowley, M. Hadjimichael, M. N. Ali, Y. C. Durmaz, J. C. Lashley, R. J. Cava and J. F. Scott, Quantum criticality in a uniaxial organic ferroelectric, J. Phys. Cond. Matt. (2014).

[4] S. E. Rowley, L. J. Spalek, R. P. Smith, M. P. M. Dean, M. Itoh, J. F. Scott, G. G. Lonzarich and S. S. Saxena, Ferroelectric Quantum Criticality, Nat. Phys. **10**, 367-372 (2014).

[5] D. E. Khmelnitskii, On low-temperature properties of uniaxial dielectrics with a soft optic mode JETP 118, 133-135 (2014).

[6] A. S. Mikheykin, E. S. Zhukova, V. I.Torgashev, *et al.*, Lattice anharmonicity and polar soft mode in ferrimagnetic M-type hexaferrite $BaFe_{12}O_{19}$ single crystal,  Eur. J. Phys. B87, Art. No. 232 (2014).

[7] T. Katsufuji and H. Takagi, Coupling between magnetism and dielectric properties in quantum paraelectric $EuTiO_3$, Phys. Rev. B 64, 054415 (2001).

[8] G. L. Tan and M. Wang, Multiferroic $PbFe_{12}O_{19}$ ceramics, J. Electroceram. 26, 170-174 (2011).

[9] G. L. Tan and M. Wang, Structure and multiferroic properties of barium hexaferrite ceramics, J. Magn. Magn. Mat. 327, 87-90 (2013).

[10] P. S. Wang and H. J. Xiang, Room-Temperature Ferrimagnet with Frustrated Antiferroelectricity: Promising Candidate Toward Multiple-State Memory , Phys. Rev. X 4, 011035 (2014).

[11] C. M. Fang, F. Kools, R. Metselaar, G. de With, R. A. de Groot, Magnetic and electronic properties of strontium hexaferrite $SrFe_{12}O_{19}$ from first-principles calculations , J. Phys. Cond. Mat. 15, 6229-6237 (2003).

[12] Shi-Peng Shen, Yi-Sheng Chai, Jun-Zhuang Cong, Pei-Jie Sun, Jun Lu, Li-Qin Yan, Shou-Guo Wang, and Young Sun, Magnetic-ion-induced displacive electric polarization in $FeO_5$ bipyramidal units of $(Ba,Sr)Fe_{12}O_{19}$ hexaferrites , Phys. Rev. B **90**, 180404R (2014).

[13] A. Hilczer, A. Bartlomiej, E. Markiewicz, et al., Effect of thermal treatment on magnetic and dielectric response of SrM hexaferrites obtained by hydrothermal synthesis, Phase Transitions 87, 938-952 (2014).

[14] E. Lieb and D. Mattis, Ordering energy levels of interacting spin systems, J. Math. Phys. 3, 749-756 (1962).

[15] L. H. Brixner, Preparation of the ferrites $BaFe_{12}O_{19}$ and $SrFe_{12}O_{19}$ in transparent form, J. Am. Chem. Soc. 81, 3841-3843 (1959).

[16] W. D. Townes, J. H. Fang, and A. J. Perrotta, Crystal structure and refinement of ferrimagnetic barium ferrite $BaFe_{12}O_{19}$, Z. Krist. Kristallgeom. Kristallphys. 125, 437-445 (1967).

[17] L. Ganapathi, J. Gopalakrishnan, and C. N. Rao, Barium hexaferrite (M-phase) exhibiting superstructure, Mat. Res. Bull. 19, 669-672 (1984).

[18] J. C. Toledano, Ferroelasticity, Annal. Telecommunications 29, 249-270 (1974).

[19] N. A. Hill, Why are there so few magnetic ferroelectrics? J. Phys. Chem. B 104, 6694-6709 (2000).

[20] N. T. M. Hien, K. Han, X.-B. Chen, J. C. Sur, I.-S. Yang, Raman studies of spin-phonon coupling in hexagonal $BaFe_{12}O_{19}$, A Raman Study of the Origin of Oxygen Defects in Hexagonal Manganite Thin Films, Chin. Phys. Lett. 29, 126103 (2012); J. Raman Spectrosc. 43, 2020-2027 (2012).

[21] X. B. Chen, N. T. M. Hien, K. Han, J. C. Sur, N. H. Sung, B. K. Cho, I. S. Yang, Raman studies of spin-phonon coupling in hexagonal $BaFe_{12}O_{19}$, J. Appl. Phys. 114, 013912 (2013).

[22] J. Muller, A. Collomb, A new representation of the bipyramidal site in the $SrFe_{12}O_{19}$ M-type hexagonal ferrite between 4.6K and 295K, J. Magn. Magn. Mater. 103, 194-203 (1992).

[23] J. F. Scott, Self-assembly and switching in ferroelectrics and multiferroics, EPL 103, 37001 (2013).

[24] J. F. Scott, Dielectric anomalies in nonferroelectric phase transitions; JETP Lett. 49, 233-235 (1989).

[25] J. Fontcuberta and X. Obradors, Dynamics of the bipyramidal ions in $SrFe_{12}O_{19}$ studied by Mossbauer-spectroscopy, J. Phys. C 21, 2335-2345 (1988).

[26] X. Obradors, A. Collomb, M. Permet, X-ray-analysis of the structural and dynamic properties of $BaFe_{12}O_{19}$ hexagonal ferrite at room-temperature, J. Sol. St. Chem. 56, 171-181 (1985).

[27] X.-B. Chen, N. T. M. Hien, K. Han, J.C. Sur, N. H. Sung, Raman studies of spin-phonon coupling in hexagonal $BaFe_{12}O_{19}$, B. K. Cho, I.-S. Yang, J. Appl. Phys. 114, 013912 (2013).

[28] R. Pirč, R. Blinc, and J. F. Scott, Mesoscopic model of a system possessing both relaxor ferroelectric and relaxor ferromagnetic properties, Phys. Rev. B **79**, 214114 (2009).

[29] J. H. Barrett, Dielectric constant in perovskite type crystals, Phys. Rev. 86, 118-120 (1952).

[30] H. B. Cao et al., High pressure floating zone growth and structural properties of ferrimagnetic quantum paraelectric $BaFe_{12}O_{19}$, APL Mater. 3, 062512 (2015).

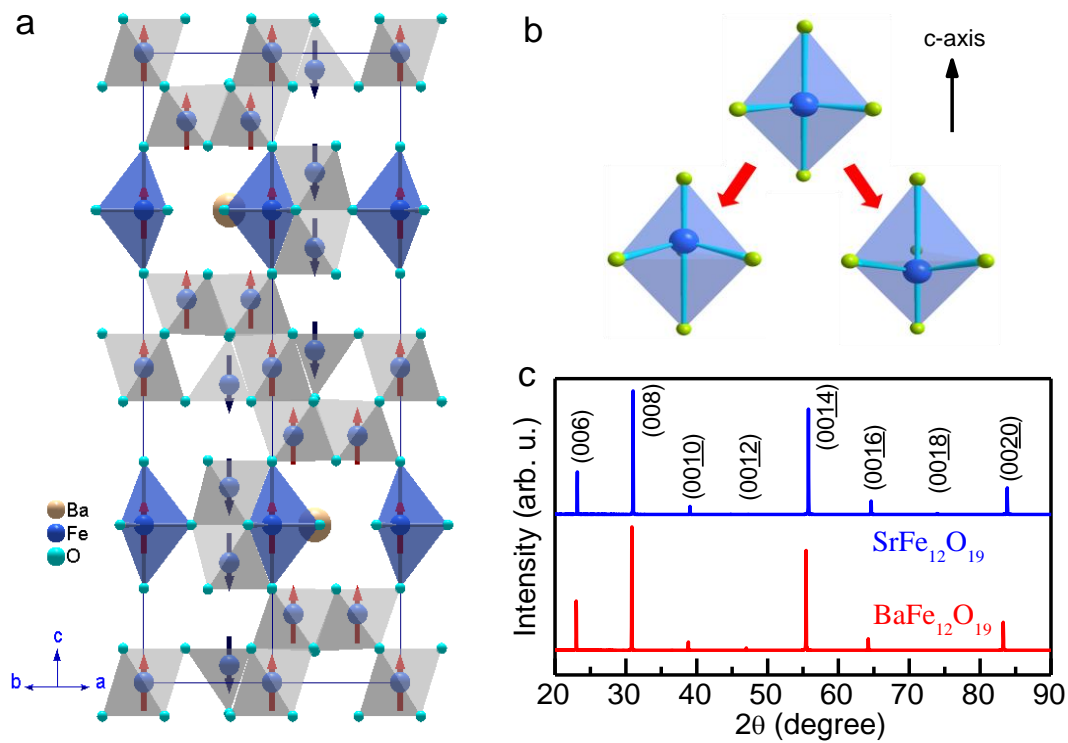

**Figure 1 - Crystal and multiferroic structures of M-type hexaferrites. a**, The crystal and magnetic structures of M-type Ba- and Sr- hexaferrites. The arrows represent the magnetic moments of $Fe^{3+}$ ions. **b**, The off-equator displacements of $Fe^{3+}$ in the $FeO_5$ bypyramidal sites induce uniaxial electric dipoles along *c* axis. Quantum fluctuations between two 4e sites prevent the onset of long-range ferroelectric ordering down to the lowest temperature. **c**, The single-crystal x-ray diffraction patterns at room temperature of prepared $BaFe_{12}O_{19}$ and $SrFe_{12}O_{19}$ crystals.

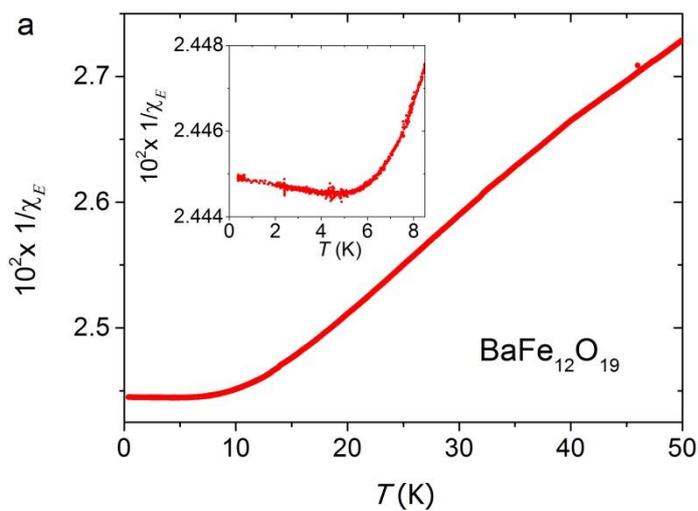

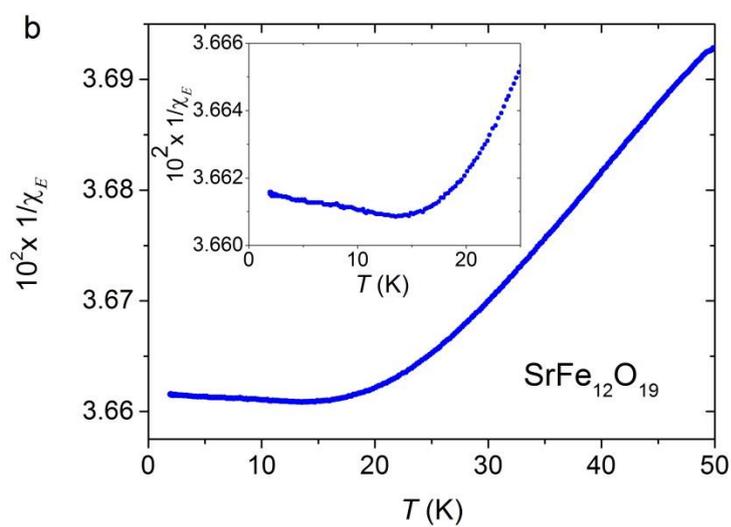

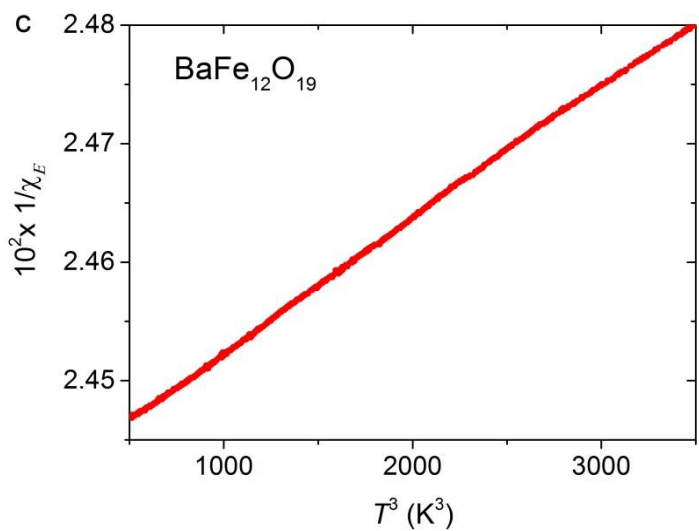

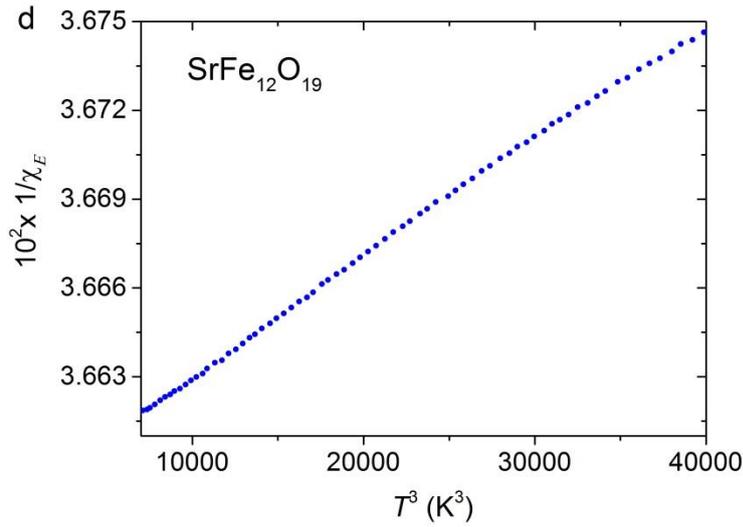

**Figure 2 - Dielectric susceptibility measurements along the c-axis in Ba and Sr M-type hexaferrites.** The main figures in (a) and (b) show the temperature depence of the inverse electric susceptibility $1/\chi_E$ below 50K for $BaFe_{12}O_{19}$ and $SrFe_{12}O_{19}$ respectively. The classical Curie-Weiss like behaviour (linear part of the curve) at higher temperaures crosses over to a different form at low temperatures due to the proximity of a ferroelectric quantum phase transition. The insets in (a) and (b) show a magnification of the low temperature region of $1/\chi_E$ against temperature in which anomolous upturns are observed. In (c) and (d) the inverse electric susceptibily is plotted against $T^3$ over the range of temperatures between the anomolous upturn at low $T$ and the classical Curie-Weiss regime at high $T$, i.e between $T_x/10$ and $T^*/10$ as explained in the text. This is between 6K and 15K for $BaFe_{12}O_{19}$ in (c) and between 20K and 35K for $SrFe_{12}O_{19}$ in (d). Recent attempts[30] by others to fit data over a wide temperature range to a single exponent are in our opinion not reliable tests.

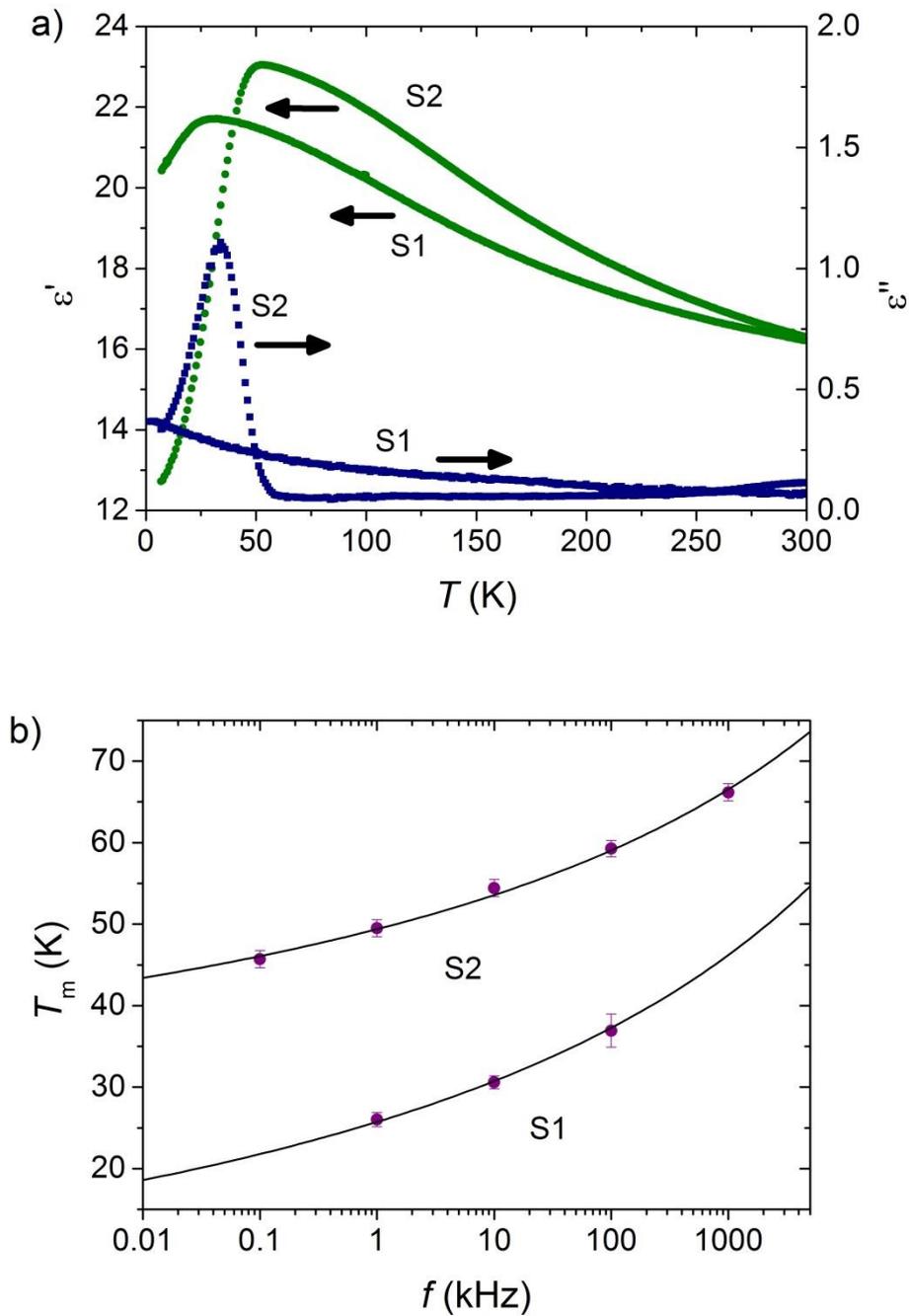

**Figure 3 - Real and imaginary parts of the dielectric constant and Vogel-Fulcher plots for $PbFe_{12-x}Ga_xO_{19}$.** Figure (a) shows the real $\varepsilon'$ and imaginary parts $\varepsilon''$ of the dielectric constant measured at 10kHz plotted against temperature $T$ for two samples of $PbFe_{12-x}Ga_xO_{19}$, S1 and S2, each with different values of x in the range 8.2 to 9.6. A Vogel-Fulcher fit to the data of the peak temperature $T_m$ versus measurement frequency $f$ is shown for the same two samples in (b). The Vogel-Fulcher equation is of the form $f = f_0 exp\left(-T_a/(T_m - T_f)\right)$ where the constants $T_a$ and $T_f$ are the activation temperature scale and freezing temperature respectively and the constant $f_0$ is a characteristic frequency. The frequency dependent variable $T_m$ is defined as the temperature at which ε' reaches its maximum value as a function of $T$, i.e. the temperature at which ε' has a peak in $T$ as in the example shown in (a). For S1 the fitting parameters were $T_a$ = 730.1K, $T_f$ = -11.66K and $f_0$ =3.016 x$10^{11}$ Hz and for S2 they were $T_a$ = 611.3K, $T_f$ = 18.07K and $f_0$ =3.001 x$10^{11}$ Hz.